\documentclass{mem}
\usepackage{natbib}\usepackage{txfonts}\usepackage{balance}
\usepackage{graphicx}
\usepackage[a4paper,breaklinks]{hyperref}
\idline{1}{1}

\usepackage{paralist}
\begin{document}
\def\teff{$T\rm_{eff }$}
\def\kms{$\mathrm {km s}^{-1}$}
\def\ea{ASA }

\title{
All-Sky-ASTROGAM -- The MeV Gamma-Ray Companion to Multimessenger  Astronomy
}


\author{
V. Tatischeff\inst{1}      
\and A. De Angelis\inst{2}     
\and M. Tavani\inst{3}
\and U. Oberlack\inst{4}  
\and R. Walter\inst{5}
\and G. Ambrosi\inst{6}
\and \\ 
A. Argan\inst{7} 
\and P. von Ballmoos\inst{8}   
\and S. Brandt\inst{9}
\and A. Bulgarelli\inst{10}  
\and A. Bykov\inst{11}
\and S. Ciprini\inst{12}
\and \\
D. Dominis Prester\inst{13}
\and V. Fioretti\inst{10}
\and I. Grenier\inst{14}
\and L. Hanlon\inst{15}
\and D. H. Hartmann\inst{16}  
\and \\ 
M. Hernanz\inst{17}
\and J. Isern\inst{17}
\and G. Kanbach\inst{18} 
\and I. Kuvvetli\inst{9} 
\and P. Laurent\inst{19}
\and M.N. Mazziotta\inst{20} 
\and \\ 
J. McEnery\inst{21}
\and S. Mereghetti\inst{22}
\and A. Meuris\inst{23}
\and A. Morselli\inst{12}  
\and K. Nakazawa\inst{24}
\and \\ 
M. Pearce\inst{25}
\and R. Rando\inst{26}  
\and J. Rico\inst{27}   
\and R. Curado da Silva\inst{28}
\and A. Ulyanov\inst{15}
\and X. Wu\inst{29}
\and \\ \vspace{-0.15cm} 
A. Zdziarski\inst{30}      
\and A. Zoglauer\inst{31}
}

\institute{ 
CSNSM, CNRS and University of Paris Sud, Orsay, France \\
\email{vincent.tatischeff@csnsm.in2p3.fr}
\and 
INFN Padova; Univ.  Udine and Padova, Italy; LIP/IST Lisboa, Portugal \\
\email{alessandro.deangelis@pd.infn.it}
\and 
INAF/IAPS, Roma, Italy
\and 
Institute of Physics and PRISMA Excellence Cluster, University Mainz, Germany
\and 
Department of Astronomy, University of Geneva, Switzerland
\and 
INFN Perugia, Italy
\and 
INAF Headquarters, Roma, Italy
\and 
IRAP Toulouse, France
\and 
DTU Space, National Space Institute, Technical University of Denmark
\and 
INAF/OAS-Bologna, Italy
\and 
Ioffe Institute, St. Petersburg, Russia
\and 
INFN  Roma ``Tor Vergata", Italy
\and
Department of Physics, University of Rijeka, Croatia
\and
AIM, CEA-IRFU/CNRS/Univ. Paris Diderot, CEA Saclay, France
\and 
School of Physics, University College Dublin, Ireland
\and 
Department of Physics and Astronomy, Clemson University, USA
\and 
Institute of Space Sciences (CSIC-IEEC), Campus UAB, Barcelona, Spain
\and 
Max-Planck-Institut fur Extraterrestrische Physik, Garching, Germany
\and 
APC, CEA/DRF CNRS, Univ. Paris Diderot, Paris, France
\and 
INFN Bari, Italy
\and 
NASA Goddard Space Flight Center, MD, USA
\and 
INAF/IASF, Milano, Italy
\and 
CEA Saclay, DRF/IRFU/DAP/LSAS, France
\and 
Department of Physics, University of Tokyo, Japan
\and 
Physics Department, Royal Institute of Technology (KTH), Stockholm, Sweden
\and 
INFN Padova, Italy
\and 
IFAE-BIST, Universitat Autonoma de Barcelona, Spain
\and 
LIP, Departamento de F\'isica, Universidade de Coimbra, Portugal
\and 
Department of Nuclear and Particle Physics, University of Geneva, Switzerland
\and 
Nicolaus Copernicus Astr. Center, Polish Academy of Sciences, Warszawa, Poland
\and 
University of California at Berkeley, Space Sciences Laboratory, USA
}
\newpage 

\authorrunning{Tatischeff et al.}

\titlerunning{All-Sky-ASTROGAM}

\abstract{All-Sky-ASTROGAM is a $\gamma$-ray observatory operating in a broad energy range, 100~keV to a few hundred MeV, recently proposed as the ``Fast'' (F) mission of the European Space Agency for a launch in 2028 to an L2 orbit. The scientific payload is composed of a unique $\gamma$-ray imaging monitor for astrophysical transients, with very large field of view (almost 4$\pi$ sr) and optimal sensitivity to detect bright and intermediate flux sources (gamma-ray bursts, active galactic nuclei, X-ray binaries, supernovae and novae) at different timescales ranging from seconds to months. The mission will operate in a maturing gravitational wave and multi-messenger epoch, opening up new and exciting synergies. 
}
\maketitle{}

\section{Introduction}

\begin{figure}[t]
\centering
\includegraphics[width=1.0\linewidth]{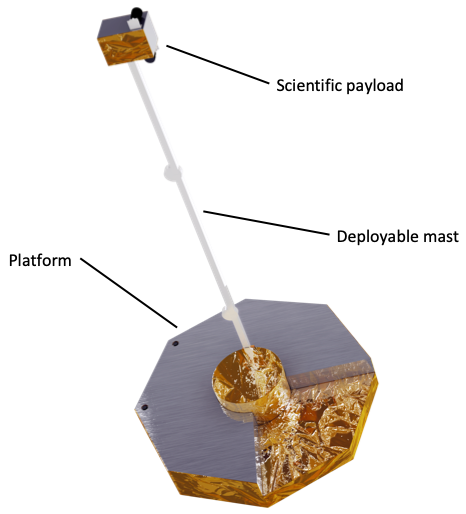}
\caption{The All-Sky-ASTROGAM satellite.
\label{fig:detsc}}
\end{figure}

In the era of multimessenger astronomy, recently opened by the discovery of simultaneous gravitational wave-$\gamma$ ray and neutrino-$\gamma$ ray signals, it is of paramount importance to have a monitor capable of detecting energetic transients in the energy range from 0.1 MeV to a few hundred MeV, with good imaging capabilities \citep[see][]{eascience}. 
In response to ESA's Call for the F mission, to be launched in 2028 with the ARIEL M4 Mission to an L2 orbit, we have proposed All-Sky-ASTROGAM (hereafter ASA), a mission dedicated to fast detection, localization, and $\gamma$-ray spectroscopy of flaring and merging activity of stellar and compact objects in the Universe, with excellent polarimetric capabilities. The instrument is a $\gamma$-ray imager covering almost $4 \pi$~sr attached to a deployable boom (Fig. \ref{fig:detsc}), capable of continuously observing every single $\gamma$-ray source in the sky. It is based on the ASTROGAM concept \citep[see][]{ea}: a telescope made of thin silicon tracking planes and a scintillation calorimeter to image the Compton interaction and the pair production by $\gamma$-rays. 


\section{Scientific goals of the mission}


\ea should make pioneering observations of the most powerful sources in the Universe, elucidating the nature of their outflows and  their effects on the surroundings. 

\begin{figure*}[ht]
\centering
\includegraphics[width=0.66\textwidth]{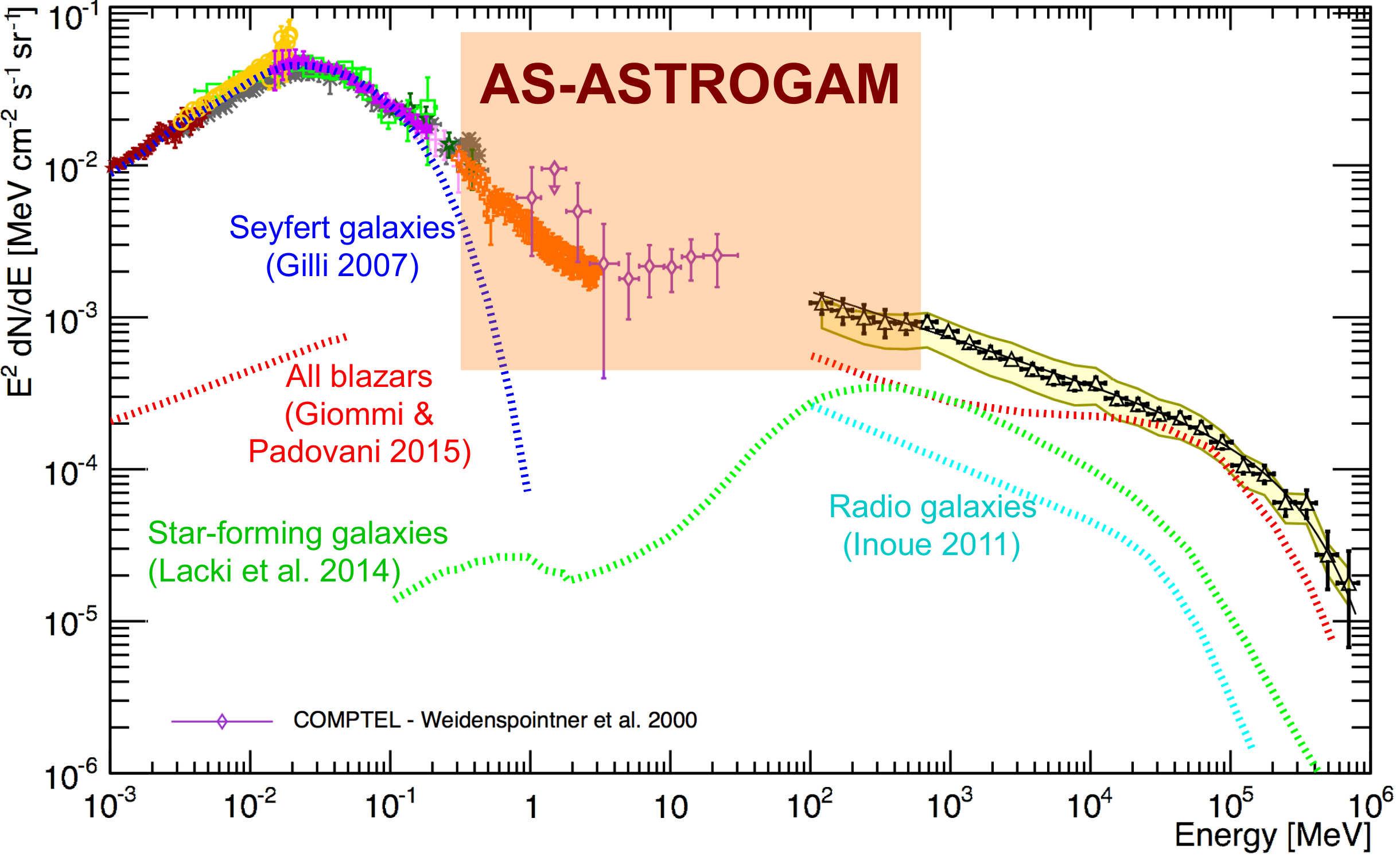}
\caption{Compilation of measurements of the total extragalactic $\gamma$-ray intensity between 1~keV and 820~GeV \citep{FermiEGB}, with different components from current models; the contribution from MeV blazars is largely unknown. The semi-transparent band indicates the energy region in which All-Sky-ASTROGAM will dramatically improve on present knowledge.
\label{fig:egb}}
\end{figure*}

\subsection{The most powerful accelerators}

Hard X-ray surveys \citep[e.g.,][]{ajello09} have been shown to be more effective in detecting \textbf{high redshift blazars} than GeV $\gamma$-ray surveys, despite the {\it Fermi} mission sensitivity, as the spectral energy distribution (SED) of these sources peak in the MeV region. \ea will detect hundreds of MeV blazars, constraining their SED peaks very tightly. These discoveries will help our understanding of (\textit{i}) the evolution of the radio-quiet and radio-loud populations of active galactic nuclei (AGN) with redshift; (\textit{ii}) the formation and growth of supermassive black holes (BHs); (\textit{iii}) the connection between the jet and the central engine and (\textit{iv}) the role of the jet in the feedback occurring in the host galaxies. \ea will substantially advance our knowledge of MeV blazars up to redshift $\sim$~3.5, with strong implications for blazar physics and cosmology. By detecting the MeV blazar population up to that redshift, the proposed mission has also the potential to resolve the extragalactic $\gamma$-ray background in a still very poorly known energy domain (Fig.~\ref{fig:egb}). 

\begin{figure*}[ht]
\begin{minipage}{0.49\linewidth}
\centering
\includegraphics[scale=0.4]{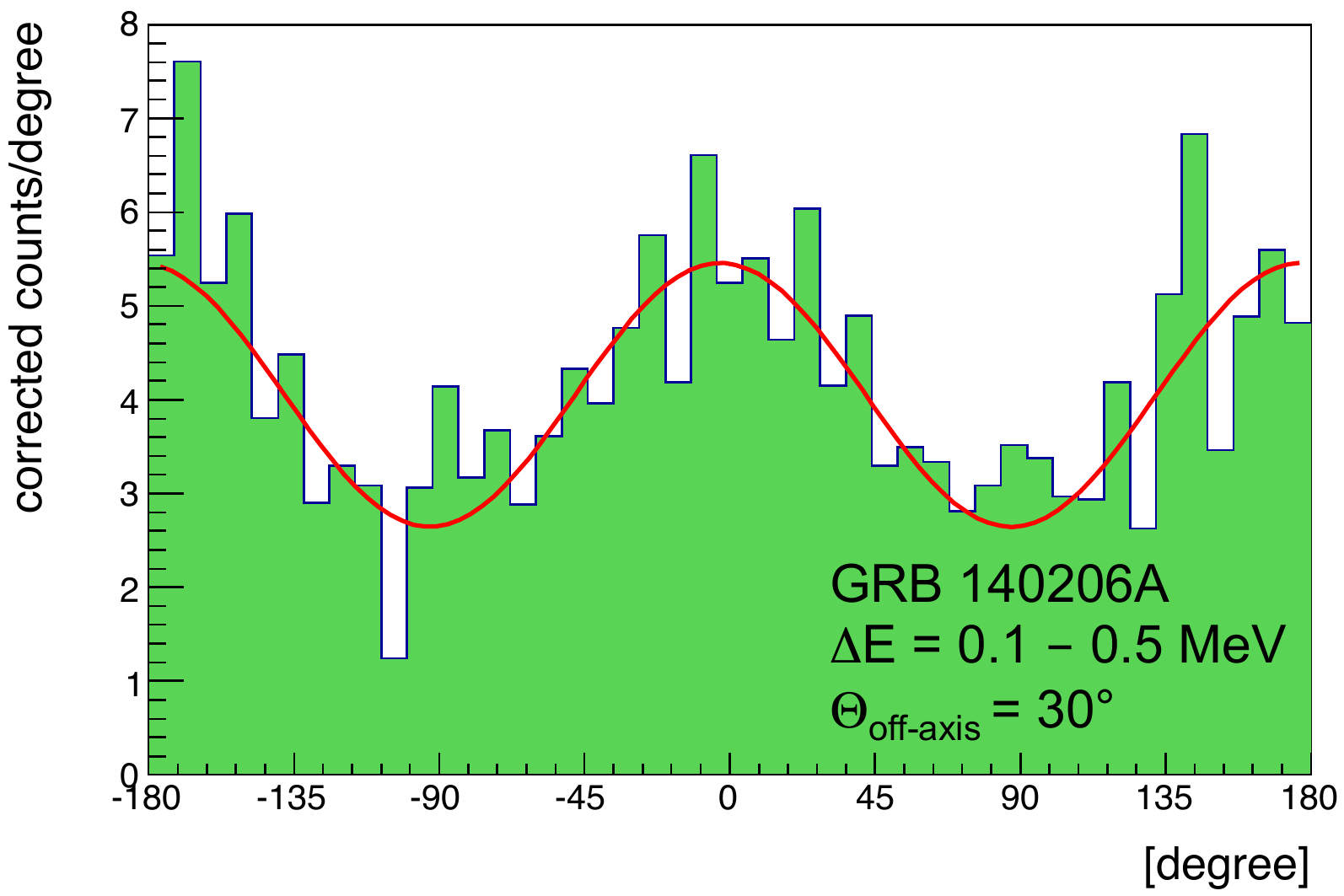}
\end{minipage}
\begin{minipage}{0.49\linewidth}
\centering
\includegraphics[scale=0.32]{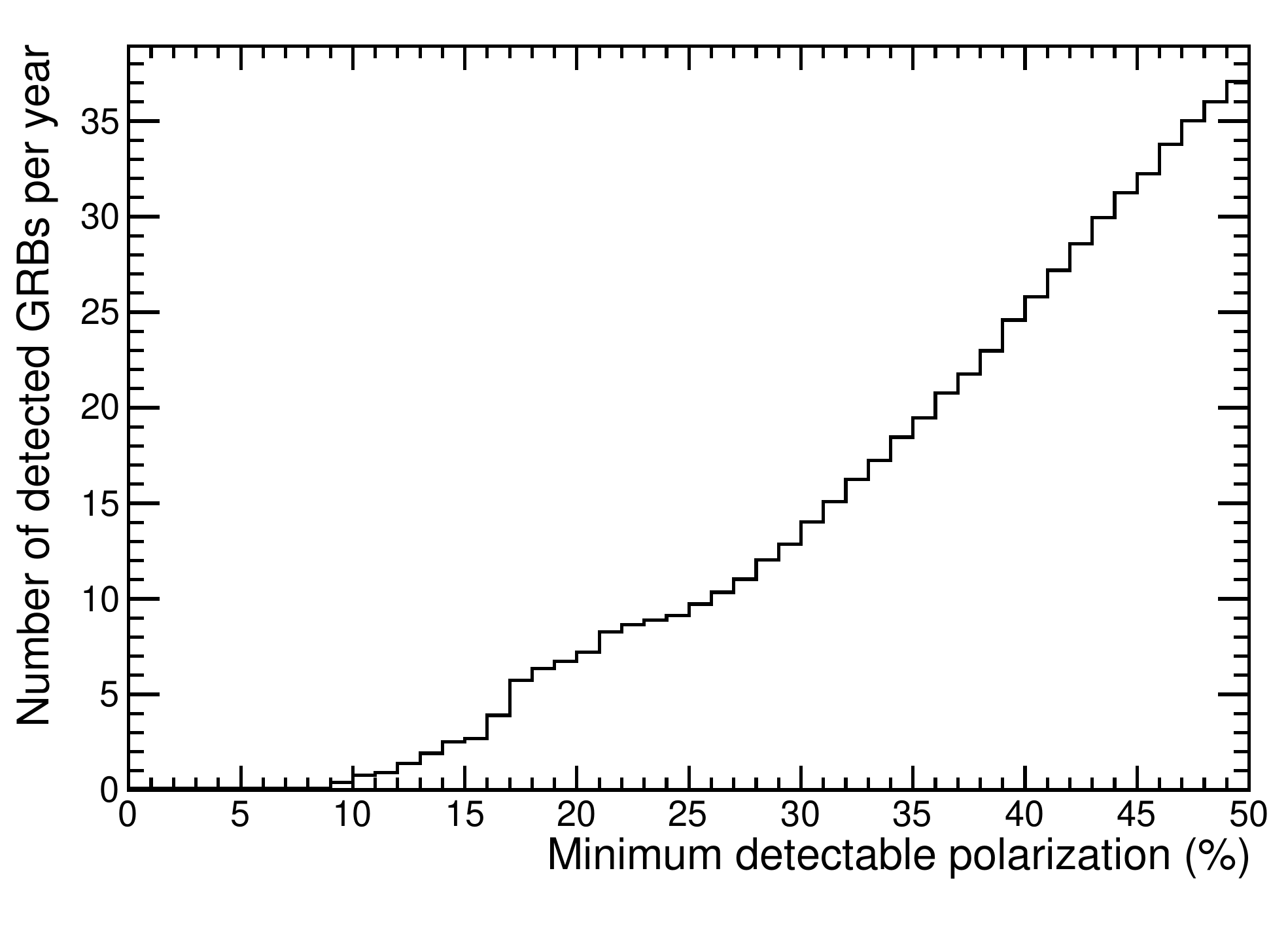}
\end{minipage}
\caption{{\it Left panel} -- All-Sky-ASTROGAM polarization response in the 100--500~keV band to a GRB like GRB~140206A ($z=2.739$, see \cite{got14}) detected at 30$^\circ$ off-axis, assuming a polarization fraction of 75\%. {\it Right panel} -- Cumulative number of GRBs to be detected by All-Sky-ASTROGAM as a function of the MDP at the 99\% confidence level.}
\label{fig:polarization}
\end{figure*}

\ea will also be an instrument studying \textbf{gamma-ray bursts} (GRBs) in one of the most important energy regions. It will be able to detect $\sim$~100 GRBs per year and accurately measure the polarization properties of several tens of GRBs per year (Fig.~\ref{fig:polarization}). This polarization signal information, combined with spectroscopy over a wide energy band, will provide unambiguous answers regarding the origin of the GRBs' highly relativistic jets and the mechanisms of energy dissipation and high-energy photon emission in these extreme astrophysical phenomena \citep[see][]{tat18}. In addition, $\gamma$-ray polarization measurement from cosmological sources provides a fundamental test of the vacuum birefringence effect and address major questions related to Lorentz Invariance Violation. 

Studying the SED of high energy emitters near compact objects is also crucial to study the interplay between accretion processes and jet emission, which can best be studied in the MeV region, where disk Comptonization is expected to fade and other non-thermal components can originate from jet particles. \ea observations of \textbf{Galactic X-ray binaries}, in particular of accreting BH systems such as Cygnus X-1, Cygnus X-3, and V404 Cygni, will determine the nature of the steady-state emission due to Comptonization and the transitions to highly non-thermal radiation. 

\subsection{Multimessenger astronomy}

The detected association between GW170817 and the GRB 170817A \citep{gravigamma} supports the connection between binary neutron star (NS) mergers and short GRBs. Joint \textbf{gravitational-wave} (GW) and electromagnetic (EM) observations are key to obtain a more complete knowledge of these sources and their environments, since they provide complementary information. From one side, GW signals provide information about the physics of the source such as, e.g., the mass and the distance; on the other hand, the identification of the possible EM counterpart pinpoints the location of the burst, possibly identifying the host galaxy and properly defining the astrophysical context. The detection of the $\gamma$-ray counterpart to GW with \ea will help to understand if also NS-BH systems are progenitors of short GRBs and characterize the astrophysical properties of the source. Current simulations predict that, after the incorporation of KAGRA and INDIGO in the GW network, between 0.2 and 6 NS-NS mergers per year will be detected by \ea in coincidence with a GW detection. 

\textbf{Neutrinos} are unique probes to study high-energy cosmic sources, since they are not deflected by magnetic fields (contrary to cosmic rays), and not absorbed by pair production via $\gamma \gamma$ interactions. The detection of cosmic neutrinos associated to EM flares from high-energy sources (e.g. AGN or microquasars) can be important for the rejection of background signals and for a better understanding of the source physics \citep[see][for the detected association between the neutrino IC170922A and the flaring blazar TXS~0506+056]{txsneutrino}, including the nature of the neutrino emission process ($p\gamma$ or $\gamma \gamma$ interactions). \ea will play a decisive role in this context thanks to its wide field-of-view, accurate sky localization, and broad energy coverage.  

\subsection{Explosive nucleosynthesis}

\ea has a very high probability to detect at least one \textbf{Type Ia supernova} (SN) in two years of nominal mission lifetime, thus allowing a direct measurement of the total mass of $^{56}$Ni/$^{56}$Co ejected by the explosion. This value is fundamental to calibrate the \cite{phi93} relation used in cosmology and the yield of synthesized Fe. Furthermore, with its all-sky coverage, \ea will be able to detect the early $\gamma$-ray emission before the maximum optical light, which is fundamental to understanding the nature of the progenitor and the explosion mechanism. Noteworthy, SNe Ia can make a significant contribution to the diffuse $\gamma$-ray background in the MeV range (Fig.~\ref{fig:egb}; see \cite{rui16}). 

Continuous monitoring of the $\gamma$-ray sky will also be essential to detect the early positron-electron annihilation emission expected before the optical maximum in classical \textbf{novae}, as well as the high-energy radiation produced through particle acceleration, in strong shocks between the ejecta and the surrounding medium –-~in recurrent symbiotic novae~-– and internal shocks in the ejecta itself –-~in classical novae. By the early detection of the high-energy $\gamma$-rays originated by neutral pions decay and/or inverse Compton scattering \citep{ferminovae}, \ea will give unique insights on the particle acceleration and mass ejection processes in novae.

The observation of the radioactive emission from the $^{44}$Ti/$^{44}$Sc chain can shed light on the clumping degree of remnants of \textbf{core-collapse SNe} as a diagnostic of internal asymmetries produced in the explosion. The sensitivity of ASA will allow the detection of this emission in most young Galactic SNRs and in the remnant of SN1987A. The determination of the amount of ejected $^{56}$Ni would also be very important to understand the physics of the explosion. 

ASA's continuous imaging of the whole sky will provide a detailed mapping of the \textbf{positron annihilation radiation} from the inner Galaxy and the diffuse $\gamma$-ray emission from the \textbf{long-lived radioactivities} $^{26}$Al and $^{60}$Fe. Building  precise maps of these Galactic emissions will shed new light on nucleosynthesis in massive stars, supernovae and novae, as well as on the structure and dynamics of the Galaxy. 

\section{Mission configuration}

\subsection{Mission profile}

The launch of the F mission to the Sun-Earth second Lagrange point offers a unique opportunity to place on a quasi-periodic orbit around L2 a sensitive $\gamma$-ray detector continuously observing every single $\gamma$-ray source in the sky during 2 years, with very good localization capabilities (e.g. 40 arcmin at 300 MeV) and excellent sensitivity to polarization in the MeV range. The $\gamma$-ray imager will be attached to a six meters-long deployable boom, which will be deployed when the spacecraft reaches its operational orbit, to reduce the $\gamma$-ray shadow cast by the satellite platform on the telescope and decrease the instrument background induced by cosmic-ray interactions with the platform materials. 

The main scientific observation mode will be the survey mode, where the spacecraft maintains a slow rotation around the boom axis while keeping the fixed solar array panels (four panels of about 1~m$^2$ each) properly oriented toward the Sun. A pointing mode will also be implemented as a fixed inertial pointing. The Attitude and Orbital Control Systems requirements are not critical, with an Absolute Pointing Error $< 1$~deg, a Relative Pointing Error $< 0.01$~deg/s and an Absolute Knowledge Error $< 30$~arcsec (to be reached after ground processing), which are to be obtained using standard class sensors and actuators. 

The ASA spacecraft platform will be made of a structure that mechanically supports the ARIEL satellite during the launch and all the spacecraft elements (boom, instrument, platform subsystems). The preliminary overall characteristics of the satellite are: 890~kg for the spacecraft wet mass and 670 W for the required power in nominal science operation. 

\subsection{Scientific payload}

\begin{figure}[t]
\centering
\includegraphics[width=0.7\linewidth]{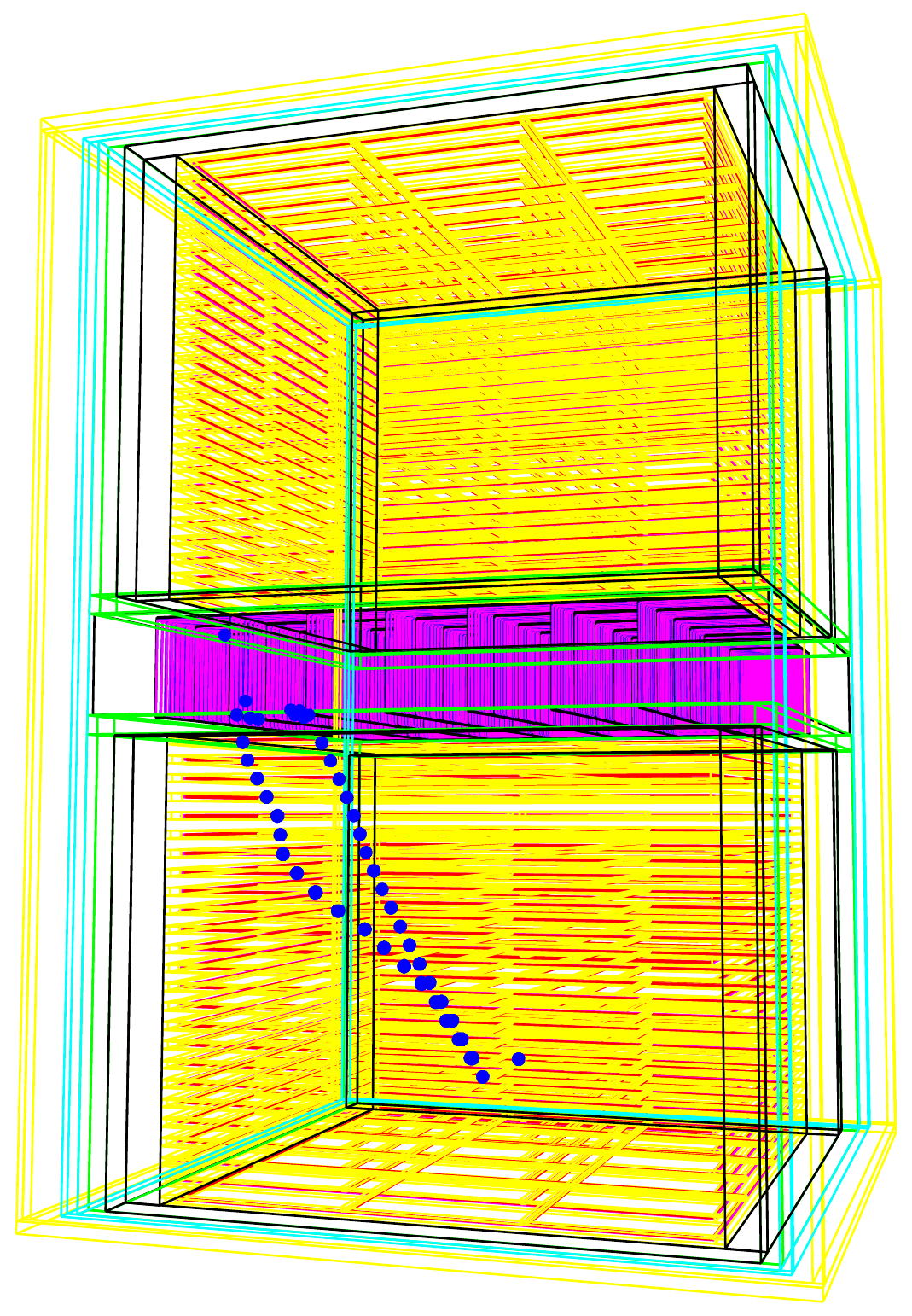}
\caption{Numerical mass model of the All-Sky-ASTROGAM imager with a simulated pair event produced by a 30-MeV photon.}
\label{fig:massmodel}       
\end{figure}

The payload of the \ea satellite consists of a single $\gamma$-ray imager operating over nearly four orders of magnitude in energy (from 100 keV to about 1~GeV) by the joint detection of photons in both the Compton ($\lesssim 10$~MeV) and pair ($\gtrsim 10$~MeV) energy ranges. The instrument is made up of three detection systems \citep[see][]{ea}: a silicon Tracker in which the cosmic $\gamma$-rays undergo a Compton scattering or a pair conversion, a Calorimeter to absorb and measure the energy of the scattered $\gamma$-rays and secondary particles, and an anticoincidence (AC) system to veto the prompt-reaction background induced by charged particles. The Calorimeter is placed in the middle of the instrument to allow the detection of $\gamma$-rays from almost any direction (see Fig.~\ref{fig:massmodel}). 

The \textbf{Si Tracker} comprises 450 double-sided strip detectors (DSSDs) arranged in 2$\times$25 layers of 3$\times$3 DSSDs, the 9 detectors of a layer being wire bonded strip to strip to form a 2-D ladder. Each DSSD has a geometric area of 9.5$\times$9.5 cm$^2$, a thickness of 500~$\mu$m, and a strip pitch of 240~$\mu$m. The DSSD signals are read out by 115,200 independent, ultra low-noise and low-power electronics channels with self-triggering capability.

The \textbf{Calorimeter} is a pixelated detector made of a high-$Z$ scintillation material -- Thallium activated Cesium Iodide -- for efficient absorption of Compton scattered $\gamma$-rays and electron-positron pairs. It consists of an array of 784 parallelepiped bars of CsI(Tl) of 5~cm length and 10$\times$10~mm$^2$ cross section, read out by silicon drift detectors (SDDs). The depth of interaction in each crystal is measured from the difference of recorded scintillation signals at both ends. The SDD boards with the analog readout ASICs in contact of the collecting anodes are connected via Kapton foils to the ADC ASICs, which are distributed over the sides of the Calorimeter. 

The \textbf{AC system} is composed of segmented panels of plastic scintillators covering the six faces of the instrument. The scintillator thickness is set to a minimum of 6~mm in order to get enough light to detect more than 99.99\% of the passing through relativistic charged particles. The plastic tiles are coupled to silicon photomultipliers (SiPM) by optical fibers, which gives the best efficiency of charged particle rejection.

The scientific payload is completed by the Tracker, Calorimeter and AC back-end electronics (BEE) modules, the Payload Data Handling Unit (PDHU) and the Power Supply Unit (PSU), which are hosted in the spacecraft platform. The PDHU is in charge of carrying out the payload internal control, scientific data processing, operative modes management, on board time management, and telemetry and telecommand management. 

The payload is based on a very high technology readiness level (TRL) for all detectors and associated electronics; it is the product of decades of instrument development by our groups and of successful space missions (AGILE, {\it Fermi}, AMS). 

\subsection{Deployable boom}

\begin{figure}[ht]
\centering
\includegraphics[width=0.87\linewidth]{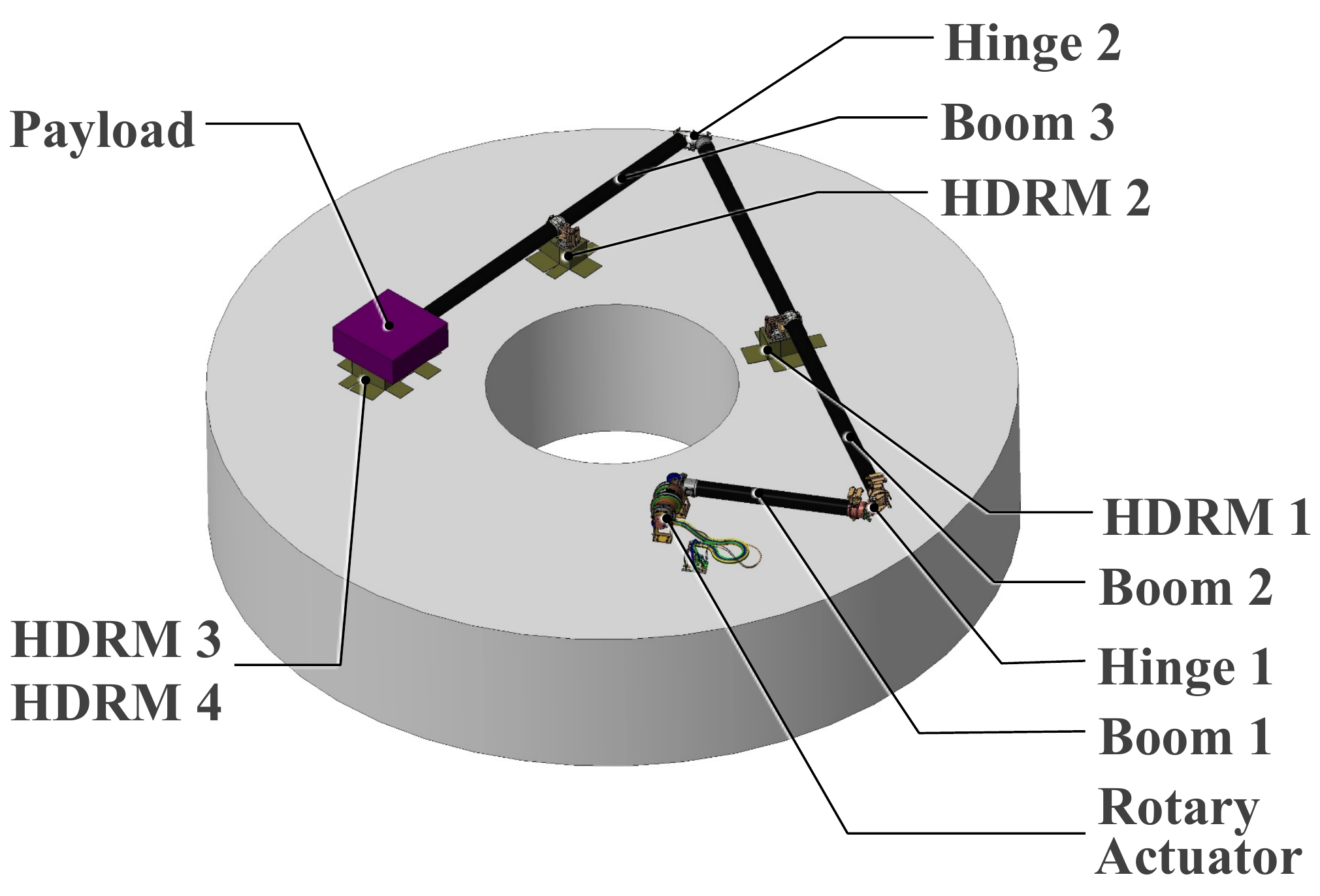}
\caption{Boom deployment mechanism.}
\label{fig:boomconcept}       
\end{figure}

The proposed concept for the deployable boom is based on an industry study performed by SENER exploiting their experience for the booms of the Solar Orbiter and JUICE Magnetometers. The deployment mechanism includes (Fig.~\ref{fig:boomconcept}): (\textit{i}) four hold-down and release mechanisms (HDRMs), (\textit{ii}) two hinge deployment mechanisms, (\textit{iii}) three booms in carbon fiber reinforced plastic, and (\textit{iv}) a rotary actuator with control electronics. 

The deployment sequence consists of three steps (see Fig.~\ref{fig:boomconcept} for the
identification of the elements): 
(\textit{i}) Hinge 2 is activated by the release of HDMRs 2, 3 and 4 to deploy Boom 3;
(\textit{ii}) Hinge 1 is activated by the release of HDMR~1 to deploy Booms 2 and 3;
(\textit{iii}) The Rotary Actuator deploys the complete assembly. 
All elements used for the  deployment have high TRL due to previous flight experience.  

\section{Conclusions}

Thanks to its continuous monitoring of the $\gamma$-ray sky with unprecedented sensitivity in the MeV range, \ea is expected to play a major role in the development of time-domain astronomy and provide valuable information for the localization and identification of gravitational-wave and high-energy-neutrino sources. 
 
\begin{acknowledgements}
We acknowledge the support of OHB Milano and SENER Madrid in the preparation of the F-Mission proposal, as well as financial support from the EU's Horizon 2020 programme under the AHEAD project (grant no. 654215).
\end{acknowledgements}

\bibliographystyle{aa}

\end{document}